\documentclass[12pt,a4paper]{JHEP3}
\usepackage{amssymb}
\usepackage{euscript}

\def\a{{\alpha}}
\def\b{{\beta}}

\def\k{\kappa}
\def\e{\textrm{e}}
\def\bra#1{\langle #1 |}
\def\ket#1{|#1 \rangle}
\def\0{\nonumber}

\def\log{{\rm log}}

\def\exp{{\rm exp}}

\newcommand\Q{{\cal{Q}}}

\newcommand\N{{\cal{N}}}

\newcommand\ee{\end{eqnarray}}      %eqnarray
\newcommand\be{\begin{eqnarray}}
\newcommand\ba{\begin{array}}           %array
\newcommand\ea{\end{array}}
\newcommand\eeq{\end{equation}}     %eqnarray
\newcommand\beq{\begin{equation}}

\preprint{SISSA--47/05/EP\\\tt hep-th/0506213}

\title{Chan-Paton factors and Higgsing from Vacuum String Field Theory}
\author{ Carlo Maccaferri\\
International School for Advanced Studies (SISSA/ISAS)\\
Via Beirut 2--4, 34014 Trieste, Italy, and INFN, Sezione di
Trieste\\
E-mail:  \email{maccafer@sissa.it} }

\abstract{We give a description of open strings stretched
between $N$ parallel D--branes in VSFT. We
 show how higgsing is generated as the branes are displaced: the shift in the mass formula for on--shell states
stretched between different branes is due to a twist anomaly, a contribution localized at the midpoint.}

\begin{document}

%%%%%%%%%%%%%%%%%%%%%%%%%%%%%%%%%%%%%%%%%%%%%%%%%%%%%%%%
\section{Introduction}
%%%%%%%%%%%%%%%%%%%%%%%%%%%%%%%%%%%%%%%%%%%%%%%%%%%%%%%%
Open String Field Theory (OSFT), \cite{Witten},  is a candidate for a
non--perturbative definition of  string theory. Central to it
there is the concept of background independence: physics should be
independent on the vacuum we decide to expand the theory on. Of
course there can be vacua on which the theory might look simpler
than in any other vacuum therefore, once such a vacuum is
identified, the theory should take its simplest form around it. As
the dynamical degrees of freedom of OSFT are open strings, a class
of its vacua should contain all possible configurations of
D--branes. Bosonic D--branes are unstable due to the omnipresence
of the open string tachyon, hence they decay classically to a
vacuum with no D--branes at all. The nature of this vacuum
(tachyon vacuum) is universal, in the sense that it is independent
of the details of the various BCFT's that describe the original
configuration of branes. For this reason there is strong
expectation that OSFT on the tachyon vacuum should be able to
describe  all open string physics and,
possibly, the related closed string physics that arises
non--perturbatively  through the classical decay of the branes,
see \cite{Taylor, Senreview, Ohmori, Arefeva, tope} for review and references therein. To overcome
the difficulties in finding exact analytic solutions of OSFT, a model (Vacuum String Field Theory) has been
conjectured
 by Rastelli Sen and Zwiebach which is a proposal for OSFT at the tachyon vacuum, \cite{Ras}.
 This model highly simplifies the form of the OSFT action by replacing the usual BRST operator
with a $c$--midpoint insertion. VSFT has passed many tests.
Classical solutions relative to different branes have been
obtained, \cite{RSZ2, RSZ3, RSZbound, BMP1}  and the right ratios
of tensions have been reproduced \cite{oku}. The small on shell
perturbations of the solution relative to the D25--brane have been
shown to cover  the open string spectrum on a D25--brane
\cite{HKw, HK, oka1, BMP2}. Time dependent solutions interpolating from the
D25--brane vacuum to the tachyon vacuum have been proposed,
\cite{BMST1, MST}. However all these successes have to be contrasted with the
singular nature of VSFT coming from the factorization of matter
and ghost degrees of freedom. Hints on how such a singular nature
arises have been given in \cite{MT, GRSZ1}: matter ghost
factorization arises as a singular limit of a reparametrization of
the worldsheet  which shrinks the whole string to its midpoint.
For this reason all of the classical solutions of VSFT exhibit a
singular behavior at the midpoint and, as a consequence,
observables are obtained by regularization of these singularities.
Although a completely consistent regularization scheme has not
been given yet (but see \cite{oka4} for interesting developments)
the leading matter--ghost factorized form of VSFT is very powerful
and, in all the above--mentioned cases, it gives rise
unambiguously to the right physics, once midpoint singularities
are correctly regularized. In this note we give a simple description of
open strings states living on a set of $N$ D--branes. When the
branes are coincident we encounter in the spectrum $N^2$ massless
vectors, giving rise to a $U(N)$ gauge symmetry. This symmetry is
part of the huge gauge symmetry of VSFT when one considers
matter--ghost factorized gauge transformations. The Chan Paton
factors arises from particular combinations of left/right
excitations on the sliver, that takes the form the generalized
Laguerre polynomials discovered in \cite{BMS3}, see also \cite{tope}. This $U(N)$
structure is dynamically generated  (it is an intrinsic part of a
classical solution) and  there is no need to add it by hand as in
first quantized string theory or even in usual OSFT. In this
sense background independence is manifest.\\
Using the translation operator $\e^{ix\hat p}$ we construct an
array of D24--branes and analyze its small on shell fluctuations.
We show that open strings stretched between parallel branes at
different positions are obtained by translating differently the
left and right part of the classical solution. This is possible
because the lump projector is left/right factorized. Of course
this operation is ambiguous for what concerns the midpoint, since
it does not have a  left/right decomposition. Indeed we show that a
naive use of left/right orthogonality cannot give rise to the
correct shift in the mass formula, proportional to the
{\it distance}$^2$ between two D--branes. By using wedge--state
regularization we show that in the sliver limit there is a non
vanishing contribution which is completely localized at the
midpoint and gives rise to the correct shift in the mass formula.
The mechanism is that of a twist anomaly, \cite{hm}, which has
proven to be crucial for obtaining the spectrum of strings around
a single D25--brane and to give the correct ratio of D--branes.

The note is organized as follows. After a brief review  of the
simplest classical solutions, the sliver and the lump,
\cite{RSZ2}, in  section 3
 we review the construction of orthogonal Neumann ({\it i.e.} zero momentum) projectors
using half string vectors and Laguerre polynomials, given in
\cite{BMS3}. In the following section we show how open string
states with the correct Chan--Paton factors arise on a classical
solution given by the sum of $N$ such orthogonal projectors. In
section 5 we consider a superposition of D24--branes and show how
the transverse fluctuations are obtained, by simply exciting the
transverse part of the classical solution with generic
oscillators. Most of these are pure gauge but  the excitation
given by the midpoint oscillator cannot be gauged away. In section
6 we higgs the above system by translating  the D24--branes at a
certain distance  from one another, thus creating an array in the
transverse direction. We analyze the fluctuations of such a
classical solution and show that the mass shift for strings
stretched between different branes is generated by a twist
anomaly.  We further show that the wave--functional for such
states is not continuous at the midpoint but gives rise to the
expected change in the boundary conditions, living the midpoint
position in the target space undetermined.

  Some computations involving well known formulas for computing $*$--products
  on Neumann--type solutions are summarized in the appendix.
%%%%%%%%%%%%%%%%%%%%%%%%%%%%%%%%%%%%%%%%%%%%%%%%%%%%%%%
\section{D--branes as projectors}
%%%%%%%%%%%%%%%%%%%%%%%%%%%%%%%%%%%%%%%%%%%%%%%%%%%
The leading order VSFT action is
\beq
{\cal S}(\Psi)= -K  \left(\frac 12 \langle\Psi
|{\cal Q}|\Psi\rangle +
\frac 13 \langle\Psi |\Psi *\Psi\rangle\right)\label{sftaction}
\eeq
where
\beq
{\cal {Q}} =  \frac1{2i}\left(c(i)-c(-i)\right)\label{calQ}
\eeq
 given the purely ghost form of the kinetic operator, the classical solutions can be matter/ghost factorized
\beq
\Psi= \Psi_m \otimes \Psi_g\label{ans}
\eeq
so the equation of motions factorizes too
\be
{\cal Q} \Psi_g & = & - \Psi_g *_g \Psi_g\label{EOMg}\\
\Psi_m & = & \Psi_m *_m \Psi_m\label{EOMm}
\ee

One can thus consider a universal ghost solution and concentrate in finding projectors of the matter star algebra.
The operator definition of the $*_m$ product is
\beq
_{123}\!\langle V_3|\Psi_1\rangle_1 |\Psi_2\rangle_2 =_3\!\langle
\Psi_1*_m\Psi_2|,
\label{starm}
\eeq
where $\langle V_3|$ is the three string vertex, see \cite{GJ1, Ohta, leclair1,  tope}.
In the following we need both translationally
invariant (D25--branes) and non-translationally invariant (Dk--branes) solutions. For simplicity we will consider the sliver and the lump, \cite{RSZ2}.
 The former is translationally invariant and
 is defined by
\beq
|\Xi\rangle = \N e^{-\frac 12 a^\dagger\cdot S\cdot a^\dagger}|0\rangle,\quad\quad
a^\dagger\cdot S \cdot a^\dagger = \sum_{n,m=1}^\infty a_n^{\mu\dagger} S_{nm}
 a_m^{\nu\dagger}\eta_{\mu\nu}\label{Xi}
\eeq
where $S= CT$ and
\beq
T= \frac 1{2X} (1+X-\sqrt{(1+3X)(1-X)})\label{sliver}
\eeq
where $C$ is the twist matrix and $X=CV^{11}$ is the diagonal Neumann coefficient, see \cite{GJ1,Ohta, tope, leclair1} for details.

To represent lower dimensional brane we need to break translation invariance in the transverse direction, this is usually
done by passing to the oscillator basis
\be
a_0^{(r)\alpha} = \frac 12 \sqrt b \hat p^{(r)\alpha}
- i\frac {1}{\sqrt b} \hat x^{(r)\alpha},
\quad\quad
a_0^{(r)\alpha\dagger} = \frac 12 \sqrt b \hat p^{(r)\alpha} +
i\frac {1}{\sqrt b}\hat x^{(r)\alpha}, \label{osc}
\ee
where $b$ is a free parameter and $\alpha$, $\beta$ denote transverse indices.

The transverse part of the vertex in this new basis becomes
\be
|V_{3,\perp}\rangle'= K\, e^{-E'}|\Omega_b\rangle\label{V3'}
\ee
with
\be
K= \left(\frac {\sqrt{2\pi b^3}}{3(V_{00}+b/2)^2 }\right)^{\frac k2},\quad\quad
E'= \frac 12 \sum_{r,s=1}^3 \sum_{M,N\geq 0} a_M^{(r)\a\dagger}
V_{MN}^{'rs} a_N^{(s)\b\dagger}\eta_{\a\b}\label{E'}
\ee
where $M,N$ denote the couple of indices $\{0,m\}$ and $\{0,n\}$,
respectively.
The coefficients $V_{MN}^{'rs}$ are given in Appendix B of \cite{RSZ2}.

The lump solution $|\Xi'_k\rangle$ has the form (\ref{Xi})
with $S$ along
the parallel directions and $S$ replaced by $S'$ along the perpendicular ones.
In turn $S'=CT'$ and $T'$ has the same form as $T$ eq.(\ref{sliver}) with
$X$ replaced by $X'$.

%%%%%%%%%%%%%%%%%%%%%%%%%%%%%%%%%%%%%%%%%%%%%%%%%%%%%%%
\section{$N$ coincident D25-branes}
%%%%%%%%%%%%%%%%%%%%%%%%%%%%%%%%%%%%%%%%%%%%%%%%%%%
There are several ways to construct coincident branes solutions in VSFT, the one we are going to use is
in terms of Laguerre polynomials, explicitly given in \cite{BMS3}.\\
Consider a left string vector $\xi^\mu_n$, such that
\be
\rho_R \xi^\mu&=&0\\
\rho_L \xi^\mu&=&\xi^\mu
\ee
The $\rho_{R,L}$ operators project into the the right/left Hilbert space at zero momentum, see \cite{RSZ3}.

 With this half string vector it is
possible to excite left-right symmetrically a string configuration, using the operator
\beq
{\bf x}=(a_\mu^\dagger,\xi^\mu)(a_\nu^\dagger,C\xi^\nu)=y \tilde y
\eeq
where $(\cdot,\cdot)$ means inner product in level space and the operators $\tilde y$ $y$ are
 identified with right/left excitations.
The half string vector $\xi$ is normalized by the following condition and definition
\be\label{norm}
(\xi_\mu,\frac{1}{1-T^2}\xi^\mu)&=&1\\
(\xi_\mu,\frac{T}{1-T^2}\xi^\mu)&=&-\k
\ee
where $T=CS$ is the Sliver Neumann coefficient, (\ref{sliver}).\\
For every choice of $\xi$ satisfying \ref{norm}, we can construct an infinite family of orthogonal projectors (D--branes) given by \cite{BMS3}
\beq
\ket{\Lambda_n}=(\k)^n\,L_n\left(\frac{\bf x}{\k}\right)\ket\Xi
\eeq
where $L_n(x)$ is the n-th Laguerre polynomial.
These states obey the remarkable properties
\be
\ket{\Lambda_n}*\ket{\Lambda_m}&=&\delta_{nm}\ket{\Lambda_m}\label{starort}\\
\bra{\Lambda_n}\Lambda_m\rangle&=&\delta_{nm}\bra\Xi\Xi\rangle\label{bpzort}
\ee
Due to these properties, once the sliver is identified with a single D--brane, a stack of $N$ D-branes can be given by
\beq\label{stuck}
\ket{N}=\sum_{n=0}^{N-1} \ket{\Lambda_n}
\eeq
From (\ref{bpzort}) we further get that the $bpz$ norm of such a solution is $N$--times the one of the sliver.

So far we have considered left-right symmetric projectors which are in one to one correspondence with
type 0 Laguerre polynomial, there are however non left-right symmetric states corresponding to generalized
Laguerre polynomials, they are given by, \cite{tope}
\be
\ket{\Lambda_{nm}}=\sqrt{\frac{n!}{m!}} \k^m (iy)^{n-m} L_m^{n-m}\left(\frac x\k\right)\ket\Xi\quad n\geq m\\
\ket{\Lambda_{nm}}=\sqrt{\frac{m!}{n!}} \k^n (i\tilde y)^{n-m} L_n^{m-n}\left(\frac x\k\right)\ket\Xi\quad m\geq n
\ee
and obey the properties \footnote{Another realization of this algebra is given in \cite{Furu1}}
\be\label{UNalg}
\ket{\Lambda_{nm}}*\ket{\Lambda_{pq}}=\delta_{mp}\ket{\Lambda_{nq}}\\
\bra{\Lambda_{nm}}\Lambda_{pq}\rangle=\delta_{mp}\delta_{nq}\bra\Xi\Xi\rangle
\ee
note in particular that $\ket{\Lambda_{n}}=\ket{\Lambda_{nn}}$. With these states
we can implement partial--isometry--like operations, see also \cite{Mamone}. Consider indeed
\be
\ket{-+}&=&\sum_{n=0}^{N-1}\ket{\Lambda_{n+1,n}}\\
\ket{+-}&=&\sum_{n=0}^{N-1}\ket{\Lambda_{n,n+1}}
\ee
It's trivial to see that
\be
\ket{+-}*\ket{-+}&=&\ket N\\
\ket{-+}*\ket{+-}&=&\ket N -\ket\Xi
\ee
Note that any of the previous  states can be obtained starting from the sliver by star products
\be\label{gener}
\ket{\Lambda_{nm}}=\left(\ket{-+}\right)_*^n*\ket\Xi*\left(\ket{+-}\right)_*^m
\ee
We have in particular
\be\label{killsli}
\ket{+-}*\ket\Xi&=&0\\
\ket\Xi*\ket{-+}&=&0\0
\ee
As a final remark it is worth noting that the partial isometry that relates projectors to projectors is actually a $*$--rotation and hence a
(matter ghost factorized) gauge transformation. We have indeed
\be\label{starot}
\Lambda_{nn}=\e^{\frac\pi 2(\Lambda_{nm}-\Lambda_{mn})}\, \Lambda_{mm} \, \e^{-\frac\pi 2(\Lambda_{nm}-\Lambda_{mn})}
\ee
as can be easily checked from (\ref{UNalg})

%%%%%%%%%%%%%%%%%%%%%%%%%%%%%%%%%%%%%%%%%%%%%%%%%%%%%
\section{$U(N)$ open strings}
%%%%%%%%%%%%%%%%%%%%%%%%%%%%%%%%%%%%%%%%%%%%%%%%%

Let's recall that the (matter--ghost factorized) open string cohomology around a  (matter--ghost factorized) classical solution $\ket\Psi$ is given by
the following conditions
\be
\ket\phi&=&\ket\phi*\ket\Psi+\ket\Psi*\ket\phi\\
\ket\phi&\neq&\ket\Lambda*\ket\Psi-\ket\Psi*\ket\Lambda
\ee
The first representing $\Q_\Psi$--closed states while the second gauges away $\Q_\Psi$ exact--states.\footnote{These conditions actually cover
 only the ghost--matter factorized cohomology}\\
In the case of $N$--coincident D25--branes the classical solution is given by (\ref{stuck}).

As multiple D--branes are obtained starting from the sliver by multiple $*$--products via (\ref{gener}), a generic open string state
on the sliver can acquire a Chan--Paton factor $(i,j)\in Adj[U(N)]$ in the same way.\\
Let $\ket{\{g\},p}$ be an on--shell open string state on the sliver, identified by the collection of polarization tensors $\{g\}$ and
momentum $p$. The Chan--Paton structure is simply given by
\be\label{CPstate}
\ket{(i,j);\{g\},p}=\left(\ket{-+}\right)_{*}^i*\ket{\{g\},p}*\left(\ket{+-}\right)_*^j
\ee
There is a subtlety here, related to twist anomaly, \cite{hm}, and the consequent breakdown of $*$--associativity.
Indeed the expression (\ref{CPstate}) is ambiguous in the overall normalization in front: it depends on how the various star products
involved are nested. This is so because all the states we are considering are constructed on the sliver, which fails to satisfy
unambiguously its equation of motion when states at non zero momentum enter the game. Consider for simplicity the Hata--Kawano tachyon
state, \cite{HKw, Fuchs}
\be\label{HKtach}
&&\ket{p}=\N \e^{(-ta^\dagger +ix)p}\ket{\Xi}=\N' \e^{ip\hat X\left(\frac\pi2\right)}\ket\Xi\\
&&t=3\frac{T^2-T+1}{1+T}v_0
\ee
this state satisfies (weakly) the linearized equation of motion (LEOM) with the sliver state
\be\label{G}
&\ket{p}*\ket{\Xi}=\ket{\Xi}*\ket{p}=\e^{-Gp^2}\ket{p}&\\
&G=\log2\quad\Rightarrow\quad p^2=1&
\ee
The quantity $G$ gets a non vanishing value from the region very near $k=0$ in the continuous basis, where some of the  remarkable properties,
encoding associativity, between Neumann coefficients breaks down due to singularities that are regulated in a non associative way
(like level truncation).
Indeed (\ref{G}) violates associativity if, as is the case, $G\neq0$
\be
(\ket{p}*\ket{\Xi})*\ket{\Xi}\neq\ket{p}*(\ket{\Xi}*\ket{\Xi})
\ee

 Just to fix a convention (and stressing once more that the only ambiguity is in the overall normalization)
we decide to do first all the star products at zero momentum (that do not develop twist anomaly) and, as the last operation,  multiply the result
with the state at definite momentum $\ket{\{g\},p}$

Now we show that (\ref{CPstate}) satisfies the LEOM
\be\label{leomN}
\ket{(i,j);\{g\},p}=\ket{(i,j);\{g\},p}*\ket N+\ket N*\ket{(i,j);\{g\},p}
\ee
using (\ref{killsli}) we get the relations
\be
\ket N*(\ket{-+})_*^{i}=(\ket{-+})_*^i*\ket {N-i}\\
(\ket{+-})_*^j*\ket {N}=\ket {N-j}*(\ket{+-})_*^{j}
\ee
which allow to write the LEOM as
\be
\ket{(i,j);\{g\},p}=\left(\ket{-+}\right)_*^i*{\Big(}\ket{\{g\},p}*\ket {N-j}+\ket {N-i}*\ket{\{g\},p}{\Big)}*\left(\ket{+-}\right)_*^j\0
\ee
It is proven in the appendix that
\be\label{key0}
\ket{\{g\},p}*\ket{\Lambda_{n\geq1}}=\ket{\Lambda_{n\geq1}}*\ket{\{g\},p}=0
\ee
once the following conditions on the half string vector $\xi^\mu$ are satisfied
\be
\left(t,\frac1{1\pm T}\xi^\mu\right)&=&0\label{cond1}\\
\left(g_{\mu\nu_1...\nu_n},\frac{1}{1\pm T}\xi^\mu\right)&=&0\label{cond2}
\ee
The first condition  states that the half string vector $\xi^\mu$ should be ``orthogonal'' to the on--shell tachyon vector
$t=3\frac{T^2-T+1}{1+T}v_0$, this just constrains $2D$ components of $\xi^\mu$ out of $D\infty-1$, and as such is easy to implement.
\footnote{These conditions are actually not needed if we represent the tachyon state as $\e^{ip\hat X\left(\frac\pi2\right)}\ket\Xi$ since,
up to overall normalizations, midpoint insertions commutes with the star product; the role of such conditions is to avoid  extra terms
when we use the CBH formula to pass to the oscillator expression $\e^{(-ta^\dagger +ix)p}\ket{\Xi}$.}
The second condition implies the ``orthogonality'' of the half string vector $\xi^\mu$ with any of the on--shell polarization tensors
$\{g\}$. We know from previous works, \cite{HK, BMP2}, that the level components of the polarization vectors should be identified
with the $k=0$ eigenvector(s), of the continuous basis at zero momentum, \cite{RSZ4}. Again, in order for (\ref{cond2}) to be satisfied,
it is sufficient to
ask that the components $\xi^\mu(k)$ vanishes fast enough at $k=0$, see \cite{BMP1, BMP2} for explicit realizations of this condition.\\
Given (\ref{key0}) it follows directly that
\be\label{key}
\ket{\{g\},p}*\ket {N-j}+\ket {N-i}*\ket{\{g\},p}=\ket{\{g\},p}*\ket \Xi+\ket \Xi*\ket{\{g\},p}
\ee
hence the LEOM simplifies to
\be
\ket{(i,j);\{g\},p}=\left(\ket{-+}\right)_*^i *{\Big(}\ket{\{g\},p}*\ket\Xi+\ket\Xi*\ket{\{g\},p}{\Big)}*\left(\ket{+-}\right)_*^j
\ee
This ensures the on--shellness of the state $\ket{(i,j)\;;\;\{g\},p}$ once this is true for $\ket{\{g\},p}$.
We thus recover $N^2$ kinematical copies of the spectrum on a single D--brane. Note  that the left/right structure of
these states is the same as a $U(N)$ double line notation, as the relations (\ref{UNalg}) certify. It should be noted  that
this Chan--Paton structure does not sit at the endpoints of the string, \cite{Furu2}, but is rather ``diluted'' on the string halves. This can be
traced back
to the singular field redefinition that should relate OSFT with VSFT, see the conclusions.
%%%%%%%%%%%%%%%%%%%%%%%%%%%%%%%%%%%%%%%%%%%%%%%%%
\section{$N$ coincident D24-branes}
%%%%%%%%%%%%%%%%%%%%%%%%%%%%%%%%%%%%%%%%%%%%%%%%%

A system of $N$ coincident D24--branes can be represented by\footnote{Other possibilities, for example putting the Laguerre polynomials
 on the codimension, are related to this by partial isometry and hence, due to (\ref{starot}), should be gauge equivalent}
\be\label{Nd24}
\ket N=\left(\sum_{n=0}^{N}\ket{\Lambda_n}\right)\otimes\ket{\Xi'}
\ee
where the state $\ket{\Xi'}$ is the lump solution given in \cite{RSZ2}.
The open string string sector with Lorentz indices
coming from the 25 dimensional world volume ($N^2$ tachyons,
$U(N)$--gluons, etc...) is exactly as in the previous section. In addition there are the physical states coming
from transverse excitations. These states are given by  exciting the transverse
part of the classical solution $\ket{\Xi'}$ with oscillators. The Chan--Paton degrees of freedom
are encoded in the worldvolume  part of the state as in the previous section. For example
the transverse scalars are given by\footnote{Note that once the relations (\ref{cond1}) are implemented one can recast the Chan Paton indices
directly on the classical solution and then act with oscillators to build onshell fluctuation}
\be\label{trans}
\ket{(ij);g^{25}, p_\|}=\N\; \e^{(-ta^\dagger +ix)p_\|}\ket{\Lambda_{ij}}\otimes g^{25}\cdot  a'^\dagger_{25}\ket{\Xi'}
\ee
It's easy to verify that these states satisfy the LEOM iff $ p_\|^2=0$, we have indeed
\be
\ket{(ij);g^{25}, p_\|}&=&\ket N*\ket{(ij);g^{25}, p_\|}+\ket{(ij);g^{25}, p_\|}*\ket N\\
&=&2^{-p_\|^2}\N\; \e^{(-ta^\dagger +ix)p_\|}\ket{\Lambda_{ij}}\otimes (\rho'_L+\rho'_R)g^{25}\cdot  a'^\dagger_{25}\ket{\Xi'}
\ee
The $\rho'_{L,R}$ are the left/right projectors with zero modes, see \cite{RSZ3}.

 The level vector $g^{25}$ is completely
arbitrary, but only its midpoint part is not pure gauge.
Indeed, exactly as in \cite{ima}, we can try to gauge away any of the states (\ref{trans})
\be\label{gaugetrans}
\delta\ket{(ij);g^{25}, p_\|}=\ket{Q_{ij}}*\ket N - \ket N*\ket{Q_{ij}}
\ee
where
\be\label{gaugepar}
\ket{Q_{ij}}=- \e^{(-ta^\dagger +ix)p_\|}\ket{\Lambda_{ij}}\otimes u^{25}\cdot  a'^\dagger_{25}\ket{\Xi'}
\ee
We have
\be\label{gaugekill}
\delta\ket{(ij);g^{25}, p_\|}=-\ket{(ij);(\rho'_L-\rho'_R)u^{25}, p_\|}
\ee
Thus the state is pure gauge if
\be
g^{25}=(\rho'_L-\rho'_R)u^{25}
\ee
It is well known that the operator $(\rho'_L-\rho'_R)$ just change the twist parity of a given level vector. In the diagonal basis all vectors
are paired except the one corresponding to $k=0$ that is only twist even, \cite{belov} (at least if we restrict ourselves to vectors that have
 a non vanishing
overlap with Fock--space vectors, see \cite{BMP2}). Thus the gauge transformation (\ref{gaugekill}) gauges away all the components of $g^{25}$ except
the $k=0$ one, which is the midpoint.
One can construct higher transverse excitations by applying more and more transverse oscillators as in \cite{HK, BMP2}. Again only the $k=0$
oscillator(s) are not gauge trivial.

%%%%%%%%%%%%%%%%%%%%%%%%%%%%%%%%%%%%%%%%%%%%%%%%%%
\section{Higgsing}
%%%%%%%%%%%%%%%%%%%%%%%%%%%%%%%%%%%%%%%%%%%%%%%%%%
Now we want to ``higgs'' the previous system of $N$ coincident D24--branes to an array of $N$ D24--branes, displaced of
a distance $\ell$  from one another in the transverse dimension $y$. This system is obtained by multiple
translation of the previous classical solution (\ref{Nd24}).
\be\label{array}
\ket{ N^{(\ell)}}=\sum_{n=0}^{N-1}{\Bigg(}\ket{\Lambda_n}\otimes \e^{-in\ell\hat p}\ket{\Xi'}{\Bigg)}
\ee
As in \cite{RSZ2} it is very convenient to pass to the oscillator basis by
\be
\hat p=\frac1{\sqrt b}\left(a_0+a_0^\dagger\right)
\ee
and to define the level vector
\be
\b_N=-\frac{i\ell}{\sqrt{b}}\left(1-T'\right)_{0N}
\ee
The transverse part of the n-th D24--brane in (\ref{array}) can thus be  written as
\be
\ket{\Xi'_{n}}= \e^{in\ell\hat p}\ket{\Xi'}=\e^{\frac{n^2}2 \left(\b, \frac{1}{1-T'} \b\right)+n\b\cdot \,a'^\dagger}\ket{\Xi'}
\ee
As proven in \cite{BMST2} we have
\be
\ket{\Xi'_{n}}*\ket{\Xi'_{m}}&=&\delta_{nm}\ket{\Xi'_{n}}\\
\bra{\Xi'_{n}}\Xi'_{m}\rangle&=&\delta_{nm}\bra{\Xi'}\Xi'\rangle
\ee
We recall here that the orthogonality condition comes from a divergence at $k=0$ of the continuous basis of the primed Neumann matrices.
Indeed, up to unimportant contributions,
 we have used the identification, see \cite{BMST2}
\be
\delta_{nm}=\exp\left[(n-m)^2\left(\beta,\frac{1}{1+T'}\beta\right)\right]=\exp\left[-\frac{\ell^2}{b}(n-m)^2\left(\frac{(1-T')^2}{1+T'}\right)_{00}\right]
\ee
Note that  we don't really need to use different projectors on the worldvolume as the degeneracy is lifted by the
different space--translations of the various projectors, however one can still use the $\ket{\Lambda_n}$'s in order
to maintain the orthogonality as $\ell\to0$.

Now we come to the spectrum.

Type $(n,n)$ strings (the ones stretched between the same D--brane) are obtained by translation of strings on a single
D24--brane
\be
\ket{(n,n);\{g\},p_\|}=\e^{in\ell p_\perp}\ket{\{g\},p_\|}^{(n)}
\ee
where $\ket{\{g\},p_\|}^{(n)}$ is an on--shell state of the previous section constructed on $\ket{\Lambda_{nn}}\otimes\ket{\Xi'}$.
Thus we  get $N$ copies of the spectrum of a single D24--branes: $N$ tachyons, $N$ massless vectors etc... This
gives a $U(1)^N$ gauge symmetry.

The situation changes when we want to consider strings  stretched between two different D--branes.
In this case we expect that a shift in the mass formula is generated, proportional to the square of the distance
between the two branes.
In order to construct $(i,j)$ states we have to translate the state $\ket{\Xi'}$ differently with respect
its left/right degrees of freedom.
We use the following identification for the left/right momentum
\be
\hat p&=&\hat p_L+\hat p_R\\
\hat p_{L,R}&=&\frac1{\sqrt b}\left(\rho'_{L,R}a+ \rho'_{L,R}a^\dagger\right)_0
\ee
We then consider the state
\be
\e^{-in\ell\hat p_L-im\ell\hat p_R}\ket{\Xi'}\propto\e^{(n\b_L+m\b_R)\cdot \,a'^\dagger}\ket{\Xi'}=\ket{\Xi'_{nm}}
\ee
where we have defined
\be
\b_{L,R}=\rho'_{L,R}\b
\ee
The $\rho'$ projectors obey the following properties up to midpoint subtleties, see later
\be
\rho'_L+\rho'_R=1\\
(\rho'_{L,R})^2=\rho'_{L,R}\\
\rho'_{L,R}\rho'_{R,L}=0
\ee
If we naively use these properties, using the formulas of \cite{RSZ2}, it is easy to prove that
\be
\ket{\Xi'_{nm}}*\ket{\Xi'_{pq}}=\e^{-\frac12\left((nm+pq-nq)\b\frac1{1-T'}\b+(mp-nm-pq+nq)\b\frac1{1-T'^2}\b\right)}\ket{\Xi'_{nq}}
\ee
We can then normalize the above states in order to have
\be\label{naivealg}
\ket{\hat\Xi'_{nm}}*\ket{\hat\Xi'_{pq}}=\delta_{mp}\ket{\hat\Xi'_{nq}}
\ee
where
\be
\ket{\hat\Xi'_{nm}}=\e^{\frac14 \b\frac{n^2+m^2+2nmT'}{1-T'^2}\b}\e^{(n\b_L+m\b_R)\cdot \,a'^\dagger}\ket{\Xi'}
\ee
Note that this normalization is quite formal as the quantity $\b\frac1{1+T}\b$ is actually divergent, this is not however a real problem
as open string states are not normalized by the LEOM's, moreover it should be noted that even if these  left/right non--symmetric
states have a  vanishing normalization, they give rise to non vanishing objects (the projectors) by $*$ product. \\
Consider now, for simplicity, the ``tachyon'' state stretched from the i-th brane to the j-th.
The corresponding  state  is given by
\be
\ket{(ij);p_\|}=\N\e^{(-ta^\dagger+ix)p_\|}\ket\Xi\otimes\ket{\hat\Xi'_{ij}}
\ee
Using (\ref{naivealg}) it is easy to see that the above state satisfies the LEOM
\be
\ket{(ij)\;;\;p_\|}=\ket{(ij)\;;\;p_\|}*\ket{ N^{(\ell)}}+\ket{ N^{(\ell)}}*\ket{(ij)\;;\;p_\|}=2^{-p_\|^2+1}\ket{(nm)\;;\;p_\|}
\ee
with $p_\|^2=1$, that is we don't get the usual mass shift proportional to the {\it distance}$^2$ between the two D--branes.\\
However the algebra (\ref{naivealg}) is not quite correct. To elucidate this point it is worth considering the components of the
level vector $\b$ in the continuous part of the diagonal basis of the primed Neumann matrices, see \cite{belov} for details.
\footnote{There are, of course, also
the contributions from the discrete spectrum, but they are not singular for $0<b<\infty$}
We have
\be
\b(k)=-\frac{i\ell}{\sqrt b }(1+\e^{-\frac{\pi |k|}{2}})V_0(k)
\ee
where $V_0(k)$ is the zero component of the normalized eigenvector of the continuous basis, \cite{belov}
\be
V_0(k)=\sqrt{\frac{bk}{4\sinh\frac{\pi k}{2}}}\left[4+k^2\left(\Re F_c(k)-\frac b4\right)^2\right]^{-\frac12}
\ee
The $\beta$ vector is finite at $k=0$,
\be
\b(0)=-\frac{i\ell}{\sqrt{2\pi}},
\ee
hence its left/right decomposition is not well defined.
This implies  that it is not correct to consider the quantity
\be\label{amb}
\gamma=\left(\b_L, \frac1{1+T'}\b_R\right)=-\frac{\ell^2}{b}\int_{-\infty}^{\infty}
\theta(k)\theta(-k)\frac{\left(1+\e^{-\frac{\pi |k|}{2}}\right)^2}{1-\e^{-\frac{\pi |k|}{2}}}V_0(k)^2
\ee
as vanishing since it is formally indeterminate (it is ``$0\cdot\infty$'').
Assuming that $\gamma$ is  non vanishing one easily obtains that the algebra (\ref{naivealg})
gets modified to
\be\label{corralg}
\ket{\hat\Xi'_{nm}}*\ket{\hat\Xi'_{pq}}=\delta_{mp}\,\e^{\frac14\left[\left(n-p\right)^2+\left(m-q\right)^2\right]\gamma}\,\ket{\hat\Xi'_{nq}}
\ee
Taking this modification into account we obtain
\be
\ket{(nm)\;;\;p_\|}*\ket N+\ket N*\ket{(nm)\;;\;p_\|}=2^{-p_\|^2+1+\frac14(n-m)^2\frac{\gamma}{\log 2}}\ket{(nm)\;;\;p_\|}
\ee
that gives the mass formula
\be
p_\|^2=1+\frac14(n-m)^2\frac{\gamma}{\log 2}
\ee
We recall that the mass for such a state should be given by ($\alpha'=1$)
\be
p_\|^2=1-\left(\frac{\Delta y_{nm}}{2\pi}\right)^2=1-\left(\frac{(n-m)\ell}{2\pi}\right)^2
\ee
The two formulas agrees iff
\be
\gamma=\left(\b_L, \frac1{1+T'}\b_R\right)=-\frac{\ell^2}{\pi^2}\log2
\ee
To verify this identity we need to regularize the ambiguous expression (\ref{amb}).
We do it by  substituting the lump Neumann coefficient $T'$ with the wedge states one $T'_N$. We remind that, see \cite{Furu1}
\be
T'_N&=&\frac{T'+(-T')^{N-1}}{1-(-T')^N}\\
T'_N\star T'_N&=&X' +(X'_{+},X'_{-})({1}-
{\cal T'}_N{\cal M'})^{-1}{\cal T'}_N
\left(\matrix{X'_{-}\cr X'_{+}}\right)=T'_{2N-1}
\ee

We have\footnote{Note that the $T'$ in the denominator of (\ref{amb}) is actually obtained by the projector equation $T'\star T'=T'$ that is
violated in  wedge--state regularization}
\be
\gamma=\left(\b,\rho_L(T') \frac1{1+T'\star T'}\rho_R(T')\b\right)=\lim_{N\to\infty}\left(\b\rho_L(T'_N) \frac1{1+T'_{2N}}\rho_R(T'_N)\b\right)
\ee

where we have used the $*$--multiplication between wedge states
\be
T'_N\star T'_N=T'_{2N-1}\backsimeq T'_{2N},\quad N\gg 1
\ee
The matrices $T'_N$ gets contributions from the continuous and the discrete spectrum but only the continuous spectrum is relevant
in the large $N$ limit, moreover it is only the region infinitesimally near the point $k=0$  that really contributes.
We have
\be
\gamma=\left(\frac{i\ell}{\sqrt{2\pi}} \right)^2\lim_{N\to\infty}\int_{-\infty}^{\infty}dk \rho_L^{(N)}(k)\frac{1}{1+t_{2N}(k)}\rho_R^{(N)}(k)
\ee
with
\be
t_N(k)=\frac{-\e^{-\frac{\pi k}{2}}+\left(\e^{-\frac{\pi k}{2}}\right)^{N-1}}{1-\left(\e^{-\frac{\pi k}{2}}\right)^N}\\
\rho_L^{(N)}(k)=1-\frac{1}{1+\left(\e^{-\frac{\pi k}{2}}\right)^{N-1}}\\
\rho_R^{(N)}(k)=\frac{1}{1+\left(\e^{-\frac{\pi k}{2}}\right)^{N-1}}
\ee
where we have used the expression of $\rho'_{L,R}$ in terms of the sliver matrix and the $*$ Neumann coefficients,
\cite{RSZ3}, and their (continuous) eigenvalues, \cite{belov}.
Let's evaluate the integral in the large $N$ limit ($x=-\frac{\pi k}{2},\;y=Nx$)
\be\label{twistok}
&&\int_{-\infty}^{\infty}dk \rho_L^{(N)}(k)\frac{1}{1+t_{2N}(k)}\rho_R^{(N)}(k)\0\\
&=&-\frac2\pi\int_{-\infty}^{\infty}dx\frac{\e^{Nx}\left(\e^{Nx}-1\right)}{\left(1-\e^x\right)\left(1+\e^{Nx}\right)\left(1+\e^{2Nx}\right)}
+O\left(\frac1N\right)\0\\
&=&-\frac2{N\pi}\int_{-\infty}^{\infty}dy\frac{\e^{y}\left(\e^{y}-1\right)}{\left(1-\e^{\frac yN}\right)\left(1+\e^{y}\right)\left(1+\e^{2y}\right)}
+O\left(\frac1N\right)\0\\
&=&\frac2{\pi}\int_{-\infty}^{\infty}\frac{dy}{y}\frac{\e^{y}\left(\e^{y}-1\right)}{\left(1+\e^{y}\right)\left(1+\e^{2y}\right)}+O\left(\frac1N\right)\0\\
&=&\frac2{\pi}\log2+O\left(\frac1N\right)
\ee
%A numerical integration shows that, see table 1
%\be\label{twistok}
%\lim_{N\to\infty}\int_{-\infty}^{\infty}dk \rho_L^{(N)}(k)\frac{1}{1+t_{2N}(k)}\rho_R^{(N)}(k)=\frac{2\log2}{\pi}=0.441271
%\ee
%\begin{table}
%\begin{center}\def\st{\vrule height 3ex width 0ex}
%\begin{tabular}{|l|l|l|l|l|l|l|l|} \hline
%$N$ & 100 & 500 & 1000  & 5000 & 10000 & 50000 &$\infty$
%\st\\[1ex]
%\hline
%$-\frac{2\pi\gamma}{\ell^2}$ & 0.445870 &  0.442180 & 0.441727 & 0.441362 & 0.441317 & 0.44128 & 0.441271
%\st\\[1ex]
%\hline
%
%\end{tabular}
%\end{center}
%\caption{Numerical evaluation of the integral (\ref{twistok}) with an extrapolation via a fitting function $f(N)=a+b/N$    }
%\end{table}
so we get the right value of $\gamma$. Note that, when $N\to\infty$,  the integrand  completely localizes at $k=0$, as claimed before.\\
As a last remark we would like to discuss about the position of  the midpoint in the transverse direction  for the states $\ket{\Xi'_{nm}}$.
We know that for the usual lump solution we have, \cite{MT}
\be
\hat X\left(\frac \pi2\right)\ket{\Xi'}=0
\ee
That is the lump functional  has support on string states in which the midpoint is constrained to live on  the worldvolume. This is
interpreted as a Dirichlet condition, see also \cite{hatanew}.
Moreover, since we have
\be
\left[\hat X\left(\frac \pi2\right),p\right]=[x_0,p]=i,
\ee
it is immediate to see that
\be
\hat X\left(\frac \pi2\right)\e^{-in\hat p\ell}\ket{\Xi'}=n\ell\,\e^{-in\hat p\ell}\ket{\Xi'}
\ee
So shifted branes undergoes a consistent change in boundary conditions.

The operator $\hat X\left(\frac \pi2\right)$ is proportional to the $k=0$ position operator, \cite{belov}
\be
\hat X\left(\frac \pi2\right)=2\sqrt{\pi} \hat x_{k=0}
\ee
It's easy to check that we have the following commutation relations
\be
\left[2\sqrt{\pi} \hat x_{k},p\right]&=&2i\sqrt{\frac{2\pi}{b}}V_0(k)\\
\left[2\sqrt{\pi} \hat x_{k},p_L\right]&=&2i\sqrt{\frac{2\pi}{b}}V_0(k)\theta(-k)\\
\left[2\sqrt{\pi} \hat x_{k},p_R\right]&=&2i\sqrt{\frac{2\pi}{b}}V_0(k)\theta(k)
\ee
That allows to write
\be
\lim_{k\to0^-}2\sqrt{\pi} \hat x_{k}\e^{-i(n\hat p_L+m\hat p_R)\ell}\ket{\Xi'}&=&n\ell\e^{-i(n\hat p_L+m\hat p_R)\ell}\ket{\Xi'}\label{midleft}\\
\lim_{k\to0^+}2\sqrt{\pi} \hat x_{k}\e^{-i(n\hat p_L+m\hat p_R)\ell}\ket{\Xi'}&=&m\ell\e^{-i(n\hat p_L+m\hat p_R)\ell}\ket{\Xi'}\label{midright}
\ee
The string functional relative to this state is not continuous at the midpoint, this is the reason why the correct mass shell  condition
comes out from a twist anomaly. In the singular representation of VSFT in which the
whole interior of a string is contracted to the midpoint, \cite{GRSZ1}, these properties reproduce the expected change in the left/right
boundary conditions, and show that the point $k=0$ naturally accounts for D--branes moduli.
%%%%%%%%%%%%%%%%%%%%%%%%%%%%%%%%%%%%%%%%%%%%%%%%%%%%%%%%%%%5
\section{Conclusions}
%%%%%%%%%%%%%%%%%%%%%%%%%%%%%%%%%%%%%%%%%%%%%%%%%%%%%%%%%%%%

In this paper we have addressed the problem of how Chan--Paton degrees of freedom  arise
on the physical excitations of multiple D--branes. For coincident D--branes we have proved
that these degrees of freedom can be encoded in the algebra of Laguerre polynomials, discovered
in \cite{BMS3}, we expect however that other (gauge equivalent) descriptions can be given.
We have further shown that if we consider an array of parallel displaced
D24--branes, the expected higgsing $U(N)\rightarrow U(1)^N$ takes place. The shift in the mass formula
is due to a twist anomaly, a contribution which is completely localized at the point $k=0$ of the
continuous spectrum of the Neumann matrices with zero modes. This is the same phenomenon,  discovered in \cite{hm}, that gives rise
to the correct mass formula for open string states on a single D25--brane. This result confirms once more that
observables in VSFT are associated to midpoint subtleties. This fact is hardly a surprise given
that the field redefinition relating OSFT on the tachyon vacuum to VSFT completely shrinks the body of a string
to its midpoint and ``resolves'' the endpoints into the left/right halves, \cite{GRSZ1}. This, we believe, is the  reason
why Chan--Paton factors are to be found in left/right excitations of VSFT classical solutions, and not in the endpoints, see \cite{Furu2}.

It would be interesting to give a BCFT description of our construction. In particular the properties (\ref{midleft}, \ref{midright})
suggest that stretched states should be given  by the insertion on the boundary of the lump surface state of a boundary changing
vertex operator, in a way similar to \cite{RSZbound}, but with one insertion more .
 Another interesting development would be to switch on interactions in order to
explicitly realize the $U(N)$ dynamical symmetry.

We hope we have given a clear picture on how some topics related to background
independence naturally emerge  from the string field algebra.
We believe this to be relevant in understanding the elusive nature of closed string states around the tachyon vacuum.

\acknowledgments
I thank Loriano Bonora for useful discussions and encouragement during the various stages of this work,
I  thank Frank Ferrari for a stimulating discussion and Ghasem Exirifard for help in integrations.
 This research was supported by the Italian MIUR
under the program ``Teoria dei Campi, Superstringhe e Gravit\`a''.

%%%%%%%%%%%%%%%%%%%%%%%%%%%%%%%%
\section*{Appendix}
%%%%%%%%%%%%%%%%%%%%%%%%%%%%%%%%%
\appendix
%\section{A collection of well--known formulae}
%%%%%%%%%%%%%%%%%%%%%%%%%%%%%%%%%%%%%%%%%%%%%%%%%%%%%%%%%%%%
\section{Proof of (4.12)}
%%%%%%%%%%%%%%%%%%%%%%%%%%%%%%%%%%%%%%%%%%%%%%%%%%%%%%%%%%%%
A general open string state on the sliver $\ket\Xi$ can be obtained differentiating  the generating state, see \cite{BMP2}
\be\label{genos}
\ket{\phi_\beta}=\e^{-(tp+\b)\cdot a^\dagger}\ket{\Xi}\e^{ipx}
\ee
where
\be
t=3\frac{T^2-T+1}{1+T}v_0
\ee
is the on--shell tachyon vector, \cite{HKw} and $\b_\mu$ is a level--Lorentz vector.\\
The  $\ket{\Lambda_n}$'s can be generated by the state ,\cite{BMS3}
\be
\ket{\Xi_\lambda}=\e^{\lambda\cdot a^\dagger}\ket\Xi
\ee
General formulas of \cite{RSZ3} and \cite{BMP2} allows to compute
\be
\ket{\phi_\beta}*\ket{\Xi_\lambda}&=&\e^{-Gp^2+A_{LR}(\b,\lambda)}\e^{-(tp+\rho_L\b-\rho_R\lambda)\cdot a^\dagger}\ket\Xi\e^{ipx}\\
\ket{\Xi_\lambda}*\ket{\phi_\beta}&=&\e^{-Gp^2+A_{RL}(\b,\lambda)}\e^{-(tp+\rho_R\b-\rho_L\lambda)\cdot a^\dagger}\ket\Xi\e^{ipx}
\ee
where
\be
A_{LR}(\b,\lambda)&=&-\frac12(\b\cdot,\frac{T}{1-T^2}\b)+(\b\cdot,\frac{\rho_R-T\rho_L}{1-T^2}C\lambda)-\frac12(\lambda\cdot,\frac{T}{1-T^2}\lambda)\\
&&+p\cdot\left((t,\frac{T}{1-T^2}\b)-(t,\frac{\rho_L-T\rho_R}{1-T^2}\lambda)+(t,\frac{\rho_R+T\rho_L}{1-T^2}\b)
        -(t,\frac{\rho_L-\rho_R}{1-T^2}\lambda)\right)\0
\ee
and
\be
A_{RL}(\b,\lambda)=A_{LR}(\b,\lambda){\Big |}_{\rho_L\to\rho_R,\;\rho_R\to\rho_L}
\ee
We can restrict the polarization $\beta_n^\mu$ to the $k=0$ component, indeed every physical excitation of the tachyon
wave function $\e^{-tp\cdot a^\dagger+ipx}\ket\Xi$ should be localized there, see \cite{HK, BMP2}.\\
Therefore is not restrictive to ask
\be
(\beta\cdot,f(T) \xi)=0
\ee
once the half string vector $\xi^\mu$ vanishes rapidly enough at $k=0$.\\
Moreover asking for (\ref{cond1}) to be satisfied, using (\ref{UNalg}), it is easy  to show that
\be\
\ket{\{g\},p}*\ket N+\ket N*\ket{\{g\},p}=\ket{\{g\},p}*\ket \Xi+\ket \Xi*\ket{\{g\},p},
\ee
as claimed

%%%%%%%%%%%%%%%%%%%%%%%%%%%%%%%%%%%%%%%%%%%%%%%%%%%%%%


\begin{thebibliography}{99}
%%%%%%%%%%%%%%%%%%%%%%%%%%%%%%%%%%%%%%%%%%%%%%%%%%%%%%%
\bibitem{Witten}
E.~Witten, {\it ``Noncommutative Geometry And String Field
Theory,''} Nucl.\ Phys.\ B {\bf 268} (1986) 253.
%%CITATION = NUPHA,B268,253;%%

\bibitem{Taylor}
  W.~Taylor and B.~Zwiebach,
{\it ``D-branes, tachyons, and string field theory,''}
  arXiv:hep-th/0311017.
  %%CITATION = HEP-TH 0311017;%%

\bibitem{Senreview}
  A.~Sen,
  {\it``Tachyon dynamics in open string theory,''}
  arXiv:hep-th/0410103.
  %%CITATION = HEP-TH 0410103;%%

\bibitem{Ohmori}
  K.~Ohmori,
  {\it``A review on tachyon condensation in open string field theories,''}
  arXiv:hep-th/0102085.
  %%CITATION = HEP-TH 0102085;%%

\bibitem{Arefeva}
  I.~Y.~Arefeva, D.~M.~Belov, A.~A.~Giryavets, A.~S.~Koshelev and P.~B.~Medvedev,
  {\it``Noncommutative field theories and (super)string field theories,''}
  arXiv:hep-th/0111208.
  %%CITATION = HEP-TH 0111208;%%

\bibitem{tope}
L.~Bonora, C.~Maccaferri, D.~Mamone and M.~Salizzoni, {\it ``Topics in string field theory,''}
arXiv:hep-th/0304270.
%%CITATION = HEP-TH 0304270;%%

\bibitem{Ras}
L.~Rastelli, A.~Sen and B.~Zwiebach, {\it ``Vacuum string field theory,''} arXiv:hep-th/0106010.
%%CITATION = HEP-TH 0106010;%%

\bibitem{RSZ2} L.Rastelli, A.Sen and B.Zwiebach, {\it ``Classical
solutions in string field theory around the tachyon vacuum"}, Adv.\ Theor.\ Math.\ Phys.\  {\bf 5}
(2002) 393 [hep-th/{0102112}].
%%CITATION = HEP-TH 0102112;%%

\bibitem{RSZ3} L.Rastelli, A.Sen and B.Zwiebach, {\it ``Half-strings,
Projectors, and Multiple D-branes in Vacuum String Field Theory"}, JHEP {\bf 0111} (2001) 035
[arXiv:hep-th/{0105058}].
%%CITATION = HEP-TH 0105058;%%

\bibitem{RSZbound}
  L.~Rastelli, A.~Sen and B.~Zwiebach,
  {\it ``Boundary CFT construction of D-branes in vacuum string field theory,''}
  JHEP {\bf 0111} (2001) 045
  [arXiv:hep-th/0105168].
  %%CITATION = HEP-TH 0105168;%%

\bibitem{BMP1}
L.~Bonora, C.~Maccaferri and P.~Prester, {\it ``Dressed sliver solutions in vacuum string field
theory,''} JHEP {\bf 0401} (2004) 038 [arXiv:hep-th/0311198.]
%%CITATION = HEP-TH 0311198;%%

\bibitem{oku}
  K.~Okuyama,
  {\it``Ratio of tensions from vacuum string field theory,''}
  JHEP {\bf 0203} (2002) 050
  [arXiv:hep-th/0201136].
  %%CITATION = HEP-TH 0201136;%%

\bibitem{HKw}
  H.~Hata and T.~Kawano,
  {\it ``Open string states around a classical solution in vacuum string field
  theory,''}
  JHEP {\bf 0111} (2001) 038
  [arXiv:hep-th/0108150].

\bibitem{HK}
  H.~Hata and H.~Kogetsu,
  {\it ``Higher level open string states from vacuum string field theory,''}
  JHEP {\bf 0209} (2002) 027
  [arXiv:hep-th/0208067].
  %%CITATION = HEP-TH 0208067;%%

\bibitem{oka1}
  Y.~Okawa,
  {\it ``Open string states and D-brane tension from vacuum string field theory,''}
  JHEP {\bf 0207} (2002) 003
  [arXiv:hep-th/0204012].
  %%CITATION = HEP-TH 0204012;%%

\bibitem{BMP2}
L.~Bonora, C.~Maccaferri and P.~Prester, {\it ``The perturbative spectrum of the dressed sliver,"}
 Phys.\ Rev.\ D {\bf 71} (2005) 026003  [arXiv:hep-th/0404154.]
%%CITATION = HEP-TH 0404154;%%


\bibitem{BMST1} L.~Bonora, C.~Maccaferri, R.J.~Scherer Santos, D.D.~Tolla
{\it ``Exact time-localized solutions in vacuum string field theory"}, Nucl.\ Phys.\ B {\bf 715} (2005) 413 [arXiv:hep-th/0409063].

\bibitem{MST} C.~Maccaferri, R.J. Scherer Santos, D.D.Tolla, {\it Time--localized Projectors in String Field Theory with E--field}
 Phys.\ Rev.\ D {\bf 71} (2005) 066007
[arXiv:hep-th/0501011].

\bibitem{MT} G.~Moore and W.~Taylor {\it ``The singular geometry of
the sliver"}, JHEP {\bf 0201} (2002) 004 [hep-th{0111069}].
%%CITATION = HEP-TH 0111069;%%


\bibitem{GRSZ1}
  D.~Gaiotto, L.~Rastelli, A.~Sen and B.~Zwiebach,
  {\it ``Ghost structure and closed strings in vacuum string field theory,''}
  Adv.\ Theor.\ Math.\ Phys.\  {\bf 6} (2003) 403
  [arXiv:hep-th/0111129].
  %%CITATION = HEP-TH 0111129;%%

\bibitem{oka4}
  N.~Drukker and Y.~Okawa,
  {\it ``Vacuum string field theory without matter-ghost factorization,''}
  arXiv:hep-th/0503068.
  %%CITATION = HEP-TH 0503068;%%

\bibitem{BMS3}
  L.~Bonora, D.~Mamone and M.~Salizzoni,
  {\it ``Vacuum string field theory ancestors of the GMS solitons,''}
  JHEP {\bf 0301} (2003) 013
  [arXiv:hep-th/0207044].
  %%CITATION = HEP-TH 0207044;%%

\bibitem{hm}
  H.~Hata and S.~Moriyama,
  {\it ``Observables as twist anomaly in vacuum string field theory,''}
  JHEP {\bf 0201} (2002) 042
  [arXiv:hep-th/0111034],
  %%CITATION = HEP-TH 0111034;%%
H.~Hata, S.~Moriyama and S.~Teraguchi,
  {\it``Exact results on twist anomaly,''}
  JHEP {\bf 0202} (2002) 036
  [arXiv:hep-th/0201177].
  %%CITATION = HEP-TH 0201177;%%

\bibitem{GJ1} D.J.Gross and A.Jevicki, {\it ``Operator Formulation
of Interacting String Field Theory"}, Nucl.Phys. {\bf B283} (1987) 1.

\bibitem{Ohta}
N.~Ohta, {\it ``Covariant Interacting String Field Theory In The Fock Space Representation,''} Phys.\
Rev.\ D {\bf 34} (1986) 3785 [Erratum-ibid.\ D {\bf 35} (1987) 2627].
%%CITATION = PHRVA,D34,3785;%%

\bibitem{leclair1} A.Leclair, M.E.Peskin, C.R.Preitschopf,
{\it ``String Field Theory on the Conformal Plane. (I) Kinematical Principles"}, Nucl.Phys. {\bf B317}
(1989) 411.

\bibitem{Furu1}
  K.~Furuuchi and K.~Okuyama,
  {\it ``Comma vertex and string field algebra,''}
  JHEP {\bf 0109} (2001) 035
  [arXiv:hep-th/0107101].


\bibitem{Mamone}
  D.~Mamone,
  {\it``Interpolating state in string field theory,''}
  arXiv:hep-th/0311204.
  %%CITATION = HEP-TH 0311204;%%

\bibitem{Fuchs}
  E.~Fuchs, M.~Kroyter and A.~Marcus,
  {\it``Continuous half-string representation of string field theory,''}
  JHEP {\bf 0311} (2003) 039
  [arXiv:hep-th/0307148].
  %%CITATION = HEP-TH 0307148;%%

\bibitem{RSZ4}  L.Rastelli, A.Sen and B.Zwiebach, {\it ``Star Algebra
Spectroscopy"}, JHEP {\bf 0203} (2002) 029 [arXiv:hep-th/0111281].

\bibitem{Furu2}
  K.~Furuuchi,
  {\it ``Non-commutative space and Chan-Paton algebra in open string field
  algebra,''}
  Nucl.\ Phys.\ B {\bf 640} (2002) 145
  [arXiv:hep-th/0202200].
  %%CITATION = HEP-TH 0202200;%%

\bibitem{ima}
  Y.~Imamura,
  {\it``Gauge transformations on a D-brane in vacuum string field theory,''}
  JHEP {\bf 0207} (2002) 042
  [arXiv:hep-th/0204031].
  %%CITATION = HEP-TH 0204031;%%

\bibitem{belov} D.M.Belov, {\it ``Diagonal Representation of Open String Star and
Moyal Product"}, [arXiv:hep-th/0204164].

\bibitem{BMST2}
  L.~Bonora, C.~Maccaferri, R.~J.~Scherer Santos and D.~D.~Tolla,
  {\it ``Fundamental strings in SFT,''}
to appear on Phys. Lett. B  arXiv:hep-th/0501111.
  %%CITATION = HEP-TH 0501111;%%

\bibitem{hatanew}
  H.~Hata and S.~Moriyama,
  {\it ``Boundary and midpoint behaviors of lump solutions in vacuum string field
  theory,''}
  arXiv:hep-th/0504184.
  %%CITATION = HEP-TH 0504184;%%


\end{thebibliography}
\end{document}